\documentclass[aps,prl,twocolumn,reprint]{revtex4-1}
\usepackage{graphics}
\usepackage{epstopdf}
\usepackage{epsfig}
\usepackage{graphicx}
\usepackage{epsf,epic}
\usepackage{color}
\usepackage{subfig}
\usepackage{marvosym }
\usepackage{amsmath}
\usepackage{booktabs}
\usepackage{multirow}
\usepackage{physics}
\usepackage{hyperref}
\usepackage{amsfonts}
\usepackage{wrapfig}
\usepackage{pstricks}
\usepackage{multirow}
\usepackage{fancyref}
\usepackage{pst-node}
\usepackage{bm}
\usepackage{dcolumn}
\newcommand{\etal}{\textit{et al.\ }}
\newcommand{\ie}{\textit{i.e.\ }}

\makeatletter
\usepackage{etoolbox} 
\appto{\appendix}{%
	\@ifstar{\def\theequation@prefix{A.}}%
	{}%
}
\makeatother
\begin{document}
\title{Buckled honeycomb group-V - \textbf{$S_6$} symmetric \textbf{$(d-2)$} higher order topological insulators}
\author{\href{https://www.santoshkumarradha.me}{Santosh Kumar Radha}}
\email{Corresponding author:srr70@case.edu}
\author{Walter R. L. Lambrecht}
\email{walter.lambrecht@case.edu}
\affiliation{Department of Physics, Case Western Reserve University, 10900 Euclid Avenue, Cleveland, OH-44106-7079}
\begin{abstract}
  Higher Order Topological Insulators (HOTI) are $d$-spatial dimensional systems featuring topologically protected gap-less states at their $(d-n)$-dimensional boundaries. With the help of \textit{ab-initio} calculations and tight binding models along with symmetry considerations we show that monolayer
  buckled honeycomb structures of group-V elements (Sb,As), which have already been synthesized, belong in this category  and have a charge fractionalization of $\frac{e}{2}$ at the corner states as well as weak topological edge states,
  all protected by their properties under the inversion operation which
  classify this system as a quadrupole topological insulator. 
\end{abstract}
\maketitle
Since the discovery of topological insulators (TIs), various exotic topological phases of matter have been discovered. A new class of such systems are higher-order topological insulators (HOTIs). Unlike regular TIs, $n^{th}$ order HOTIs have protected edge states in their corresponding $(d-n)$  dimension, where $d>n>1$.\cite{Benalcazar61,Langbehn,Song,Benalcazar17,Schindlereaat0346} Various
3D materials\cite{Schindlereaat0346,3dti,3dti2,Wang19,ashvin} and artificial systems\cite{Ezawa18b,Ezawa18,3dti4,Kariyado2017,3dti6} have recently been proposed to be HOTIs. It was recently also proposed that gap-less HOTI boundary states
in proximity to a superconductor could host Majorana
states without the need for special pairing mechanisms or magnetic fields.\cite{Yan19}
\begin{figure}[!htb]
  \includegraphics[width=9cm]{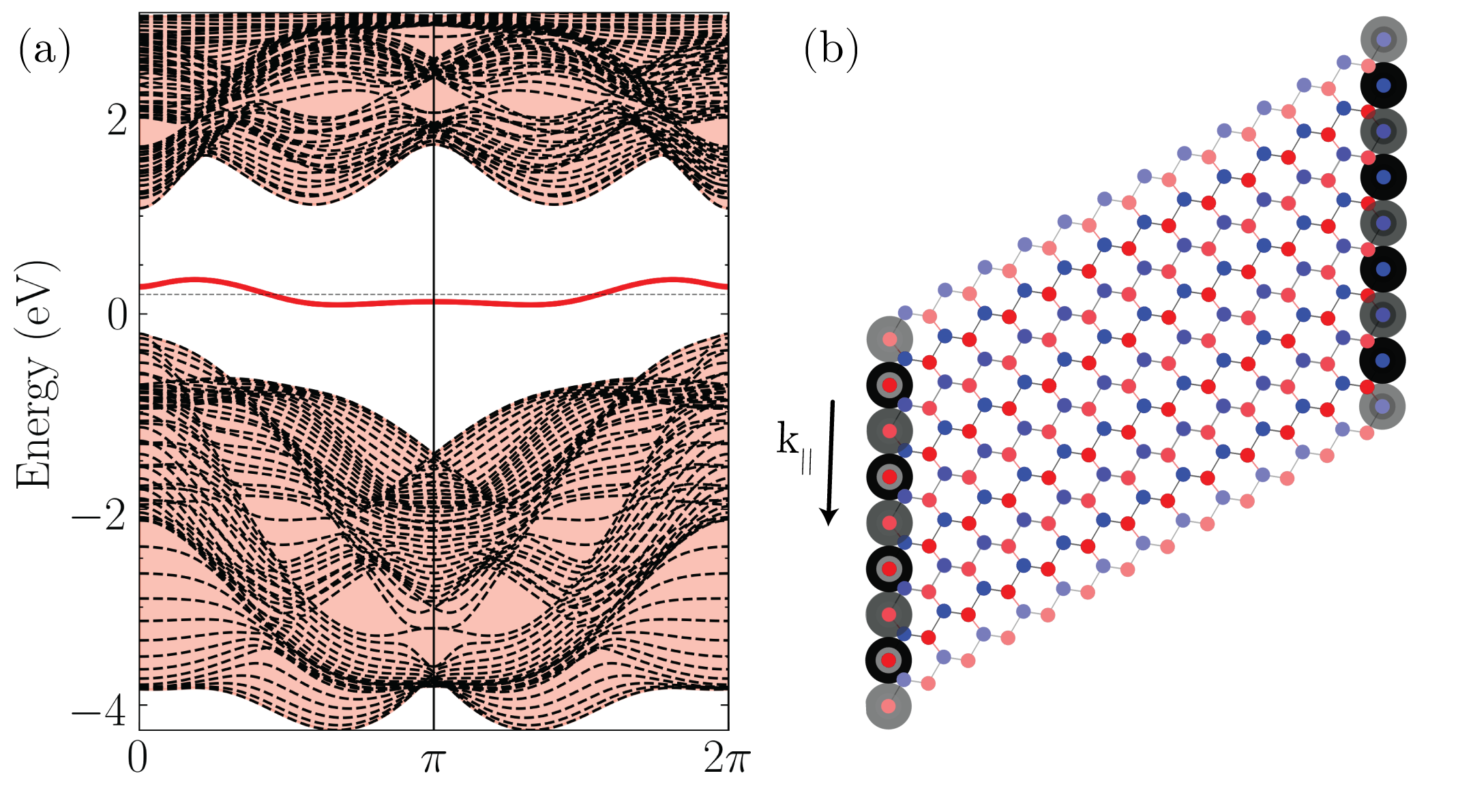} 
  \caption{(a) Density functional theory band structure  of nanoribbon
    of buckled monolayer Sb in local density approximation
    along the zigzag edge ${\bf k}_\parallel$, showing edge states(red); (b)
    structure and edge state wave function modulo squared.\label{fig:dftsurface}}. 
\end{figure}

In this letter, we show that an already existing 2D material system -
monolayer buckled honeycomb $\beta$-Sb\cite{sb_growth} is a $n=2$ HOTI that is protected by $S_6$ symmetry. Our symmetry analysis closely  follows and slightly generalizes
the model introduced by Ref\cite{Benalcazar19} for $C_n$-symmetric systems.
We will furthermore show
that it is closely related to, but not identical to, the Kekul\'e lattice,
\ie a honeycomb lattice with bond alternation. In this system both
weak edge states occur in 1D nanoribbons and corner states in 0D fragments
of the lattice and are related to the quadrupole character of the system
as defined in Refs.\onlinecite{Benalcazar61,Benalcazar17}. 

\paragraph{Background on monolayer Sb - }
In a recent  paper\cite{radha2019topological}, we studied the topological
behavior of honeycomb monolayer
Sb from the completely flat to the free-standing equilibrium buckled
structure. In that paper we showed that unlike in graphene, the $s$-states
of Sb form essentially decoupled separate bands from the $p$-orbitals
because of their energy separation in the atom. 
While in the (nearly) flat honeycomb structure, $p_z$ orbitals are also
(nearly) decoupled from the $p_x,p_y$ orbitals by the horizontal mirror plane,
the set of three $p$-orbitals in
the equilibrium buckled structure can fully interact and
form  bonding and antibonding sets of
bands separated by a gap and thus this system was until now considered
a trivial semiconductor. However, we found that relatively  flat edge states
occur in the middle of the gap of nanoribbons of the buckled
form and point to another 
type of topological effect being responsible for these surface states.
The density functional theory (DFT)
band structure results for a nanoribbon  are shown in
\autoref{fig:dftsurface} as obtained from a tight-binding model
parametrized with maximally localized Wannier function
extracted hopping integrals from a projector augmented
wave (PAW)\cite{PAW} calculation of the 2D periodic system using the Quantum
Espresso  \cite{Giannozzi2009} and Wannier90 codes.\cite{wannier90}
Computational details are given in Supplemental Material (SM).\cite{SM}

\paragraph{Qualitative discussion of origin of edge states -} 
In a broad sense the origin of these edge states is related to the
``obstructed atomic limit'' (OAL), a concept introduced by Bradlyn \etal\cite{Bradlyn2017,Cano18} Accordingly, a set of bands is in the OAL when they possess
symmetric, localized Wannier functions that reside on  Wyckoff positions
distinct from the atomic positions which cannot be smoothly deformed
to the latter. This corresponds to a weak topological phase.

\begin{figure}[!htb]
	\includegraphics[width=8cm]{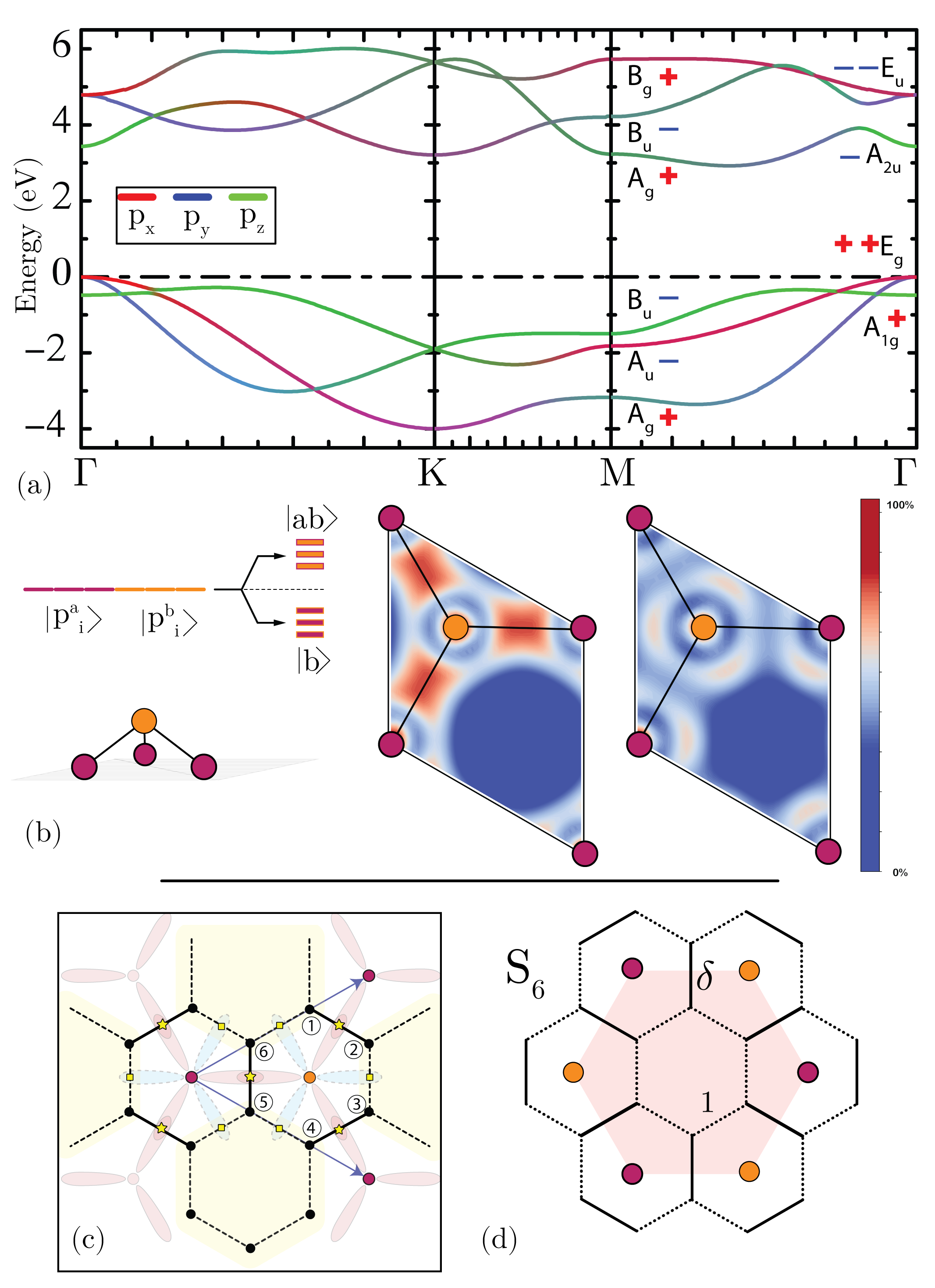} 
	\caption{\textit{(a)} QSGW band structure of honeycomb-buckled Sb;
          \textit{(b)} schematic of $p$-levels from atomic limit to
          buckled Sb and
          contour plot of total electron density from the 3 $p-$derived valence ($\ket{b}$) and conduction ($\ket{ab}$) bands; \textit{(c)} relation of
          bond centers to Kekul\'e model: see text;
          \textit{(d)} Kekul\'e model.\label{fig:bands}}
\end{figure}

In the present case,
the bands of interest result from the Sb-$\{p_x,p_y,p_z\}$ orbitals which form
three sets of bonding (occupied, valence) and three sets of antibonding
(empty,conduction)  bands.
These bands are shown in \autoref{fig:bands} along with the relevant
symmetry labeling and indicating the atomic orbital character of each band.
This band structure is actually obtained at the quasiparticle self-consistent
(QS)$GW$ level\cite{Schilfgaarde06}
which guarantees accurate band gaps but for the remainder of
this paper, this is not important and a DFT or even simpler tight-binding (TB) 
models have the same set of irreducible representations present in the valence (VB)
and conduction (CB) band manifolds as shown in SM.\cite{SM}.

Significant hybridization between all three $p$-orbitals is apparent.
To better understand the origin of the gap and hybridization, we plot
in \autoref{fig:bands}(b) the  $\sum_{n{\bf k}}|\psi_{n{\bf k}}|^2$
separately for $n\in$ VB and
CB manifolds.  These show clearly that the Wannier function
corresponding to the VB are  localized at the bond centers, while the
ones of the CB are centered at the ``antibond centers'', obtained from the
bond center by inversion about the atomic sites. This in itself is
already a clear indicator of the non-trivial nature of the band structure.
Intuitively, if one cuts the system along these bonds, dangling bond like 
edge states are expected. 
\paragraph{Relation to Kagome and Kekul\'e lattice -}
Remarkably the bond centers and antibond centers from respectively
a Kagome and \textit{anti-Kagome} lattice as illustrated in SM.\cite{SM}
More interestingly, a further transformation relates this band structure to
the Kekul\'e lattice. As shown in \autoref{fig:bands}(c)
the bond centers (yellow stars)  of the (pink) bonding orbitals of Sb atoms (orange and red circles) and the corresponding antibond (blue)
centers (yellow squares) can alternatively be viewed as the bond centers
of yet another honeycomb lattice (black circles) with alternating
bond strength. We
may recognize this as the bond modulated honeycomb lattice, known as the
Kekul\'e lattice.  We can see in \autoref{fig:bands}(d)
that this lattice consists of hexagon shaped molecules connected by (intracell) 
bonds of strength $1$ within the Wigner-Seitz unit cell and connected by
intercell bonds of strength $\delta$. It is useful to note that the bond centers
of the original buckled Sb lattice, which coincide with the bond
centers of the final Kekul\'e lattice all lie in the same horizontal
plane.  The Kekul\'e lattice is thus strictly 2D and the inversion
operator  in 2D is identical with a two-fold rotation $C_2$ about the $z$-axis.
Thus the point group of the Kekul\'e lattice is $C_6$ while that of
the Sb lattice is $S_6$ but the two are simply related by replacing
the $C_2$ operation by the inversion operation, which we will denote $\mathcal{I}$.

The topological properties of the Kekul\'e lattice have been
discussed in a number of recent papers.\cite{Kariyado2017,WuHu2016,Liu17prl,Liu17jsps,Liu19}
These papers have not only shown the presence of topological
edge states but also topological corner states in the Kekul\'e lattice
in the nontrivial condition $\delta>1$. Not surprisingly, one can think of Kekul\'e as a hexagonal 2D generalization of the Su-Schrieffer-Heeger (SSH) model\cite{SSH}
or a set of SSH models arranged next to each other. 

\begin{figure}[!htb]
	\includegraphics[width=8cm]{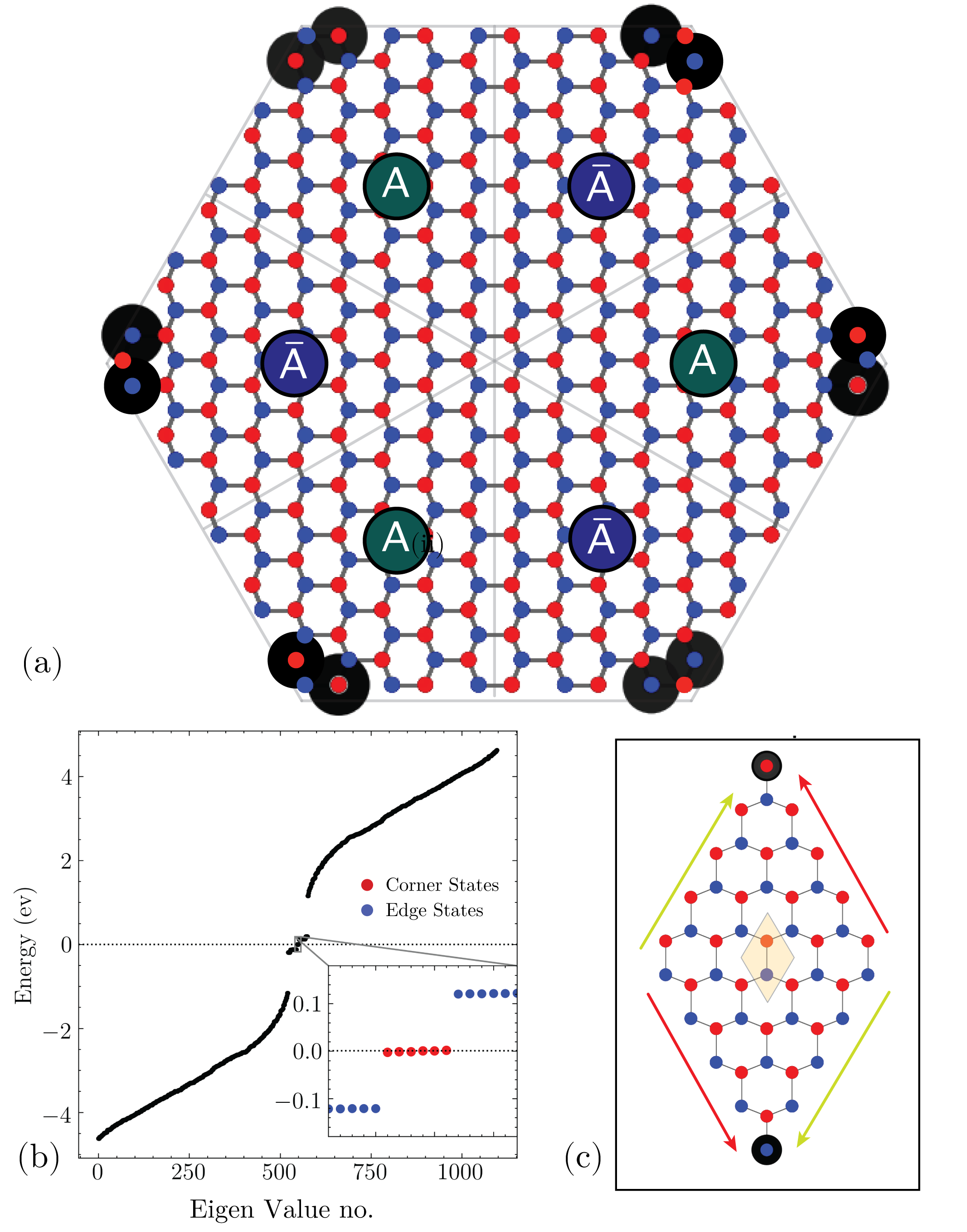} 
	\caption{(a) Finite size fragment of honeycomb (red and blue circles
          show Sb atoms below and above the plane), large black circes  
          represent $|\psi|^2$ for the localized corner states indicated in 
          in the eigenvalue spectrum in (b).\label{fig:corner}}
\end{figure}

\paragraph{Corner states -}
We now turn our attention to the higher order topologically required corner
states in a 0D system that preserves the rotational symmetries
of the periodic lattice, in other words, a finite hexagonal portion
cut out from the buckled buckled honeycomb lattice. First, let us mention
that such hexagonal nanoflakes have already been fabricated\cite{sbexp}.
Benalcazar \etal\cite{Benalcazar19} have described topological invariants
and the related occurrence of corner-states and their charge fractionalization
for $C_n$ groups.
  The conditions under which corner-states occur depend on the symmetries
  of the states at high-symmetry {\bf k}-points.  While the
  system under study here has $S_6$ instead of $C_6$ symmetry, we can
  easily generalize their procedure. According to Benalcazar \etal\cite{Benalcazar19} for the $C_6$ group, the topological invariant is
  $\chi^{(6)}=([M^{(2)}_1],[K^{(3)}_1])$,
  which, for example at $M$ indicates the difference in number
  of eigenstates in the occupied bands manifold of the $C_2$ operation
  indicated by the superscript corresponding to a given eigenvalues (1, the subscript) at $M$ and $\Gamma$. 
  A 6-fold rotation
  can be viewed as the product of a 3-fold and 2-fold rotation. Similarly,
  a 3-fold rotation-inversion is the product of a 3-fold rotation and
  an inversion. Thus we merely have to replace the two fold rotation by
  inversion for our case. Thus the important indicator becomes
  $[M^{(i)}_\pm]$ which is the difference in
  number of bands with even/odd character under the inversion operation at $M$
  and at $\Gamma$. We can see from the symmetry labeling
  in \autoref{fig:bands}(a) that the $[M^{(i)}_\pm]=\pm2$.  According
  to Table I in \onlinecite{Benalcazar19}, 
  the invariants at $K$ and $M$ are either (2,0),(0,2) or (0,0) where the
  last case is the trivial topology. Thus, our system corresponds to
  the $h^{(6)}_{3c}$ {\sl primitive generator} class in Benalcazar \etal's
  notation.
  This indeed means that there are 3 filled bands and that the generator
  is obtained from orbitals centered at the Wyckoff position $c$.
  In the buckled Sb system, the space group is $P\bar{3}c1$ or $D_{3d}^3$
  or $\#164$ and the Wyckoff position is $d$. Thus, generalizing
  their notation our primitive generator would be $h^{(\bar{3})}_{3d}$
  but in terms of charge fractionalization and polarization would have
  the same invariants as $h^{(6)}_{3c}$. (For the Kekul\'e case,
    the plane space group is $p3m1$ and the Wyckoff position is $c$.)
   It implies that there is
  no net dipole in the plane and the corner charge fractionalization will be
  $e/2$  in each $\pi/6$ sector. 

  We verify this prediction in \autoref{fig:corner}. The calculation
  is performed on a hexagonal 0D flake within the 
  $\{p_x,p_y,p_z\}$ 
    NN-TB Hamiltonian (rather than the simplified Kekul\'e model,
  which is however topologically equivalent). The eigenvalues of this Hamiltonian
  are shown in Fig. \ref{fig:corner}(b).
  One can see that near zero energy, in the band gap there occurs both
  edge states (blue dots in inset of (b)) 
  as well as  mid-gap corner states (red).  Their wave function
  modulo squared is indicated by the black circles in Fig. \ref{fig:corner}(a)
  clearly showing that
  these states are localized on the corner atoms. The $\pi/6$ sectors are
  labeled alternating $A$ and $\bar{A}$, which are related by the
  inversion symmetry. One should note that though the existence of such corner states is deeply tied to the choice of symmetric unit cell,  $\pi/6$ charge-fractionalization is not. In any general $S_6$ symmetric $0D$ fragment,
  one is guaranteed $e/2$ charge-fractionalization (other examples are
  given in SM).\cite{SM} 

\paragraph{Relation to multipole insulators -}
A closely related point of view on the origin of the corner states arises in the context
of quantized multipole insulators.
Although the net polarization in the plane is zero for our case,
the system has a non-zero quadrupole insulator character.

For a system with inversion symmetry, the contribution to the
polarization  projected on direction $i$ from band $n$ can be
obtained from the the eigenvalues of the
inversion operator at Time Reversal Invariant Momentum (TRIM) points $M$ 
 and $\Gamma$ and is given by \cite{Benalcazar17,Liu17prl,Liu19} 
\begin{equation}
P^n_i=\frac{e}{2}(q^n_i \text{ mod } 2 ) \quad \hbox{\rm with}\quad
(-1)^{q^n_i}=\frac{\eta_n(M_i)}{\eta_n(\Gamma)} \label{eq:P}
\end{equation}
where $\eta^n(\boldsymbol{k})$ is the eigenvalue of the inversion operation.
The quadrupole moment is then given by 
\begin{align}
\mathcal{Q}_{ij}=\sum_{n}^{N_{occ}} \frac{P^n_iP^n_j}{e} \label{eq:5}
\end{align}
For our system $(P^n_1,P^n_2)=(0,0),(\frac{e}{2},\frac{e}{2}),(\frac{e}{2},\frac{e}{2})$ for bands $n=1,2,3$ numbered from bottom to top. Thus the net dipole moment is trivially $(e,e)$ as it should be for a $C_6(S_6)$ symmetric system but the net quadrupole moment has magnitude $e/2$, 
the effect of which can be seen in a $d-2$ system cut
along the lattice vectors as shown in Fig. \ref{fig:corner}(c).
In this figure we see the dipoles on opposite edges canceling each other
but leading to a net quadrupole with charge accumulation at the
two corners indicated by the black circles showing $|\Psi|^2$ of the corner localized state in the TB model. 
Not surprisingly, given the close connection pointed out earlier,
similar corner states and quadrupole character are also found
for the Kekul\'e lattice in the non-trivial limit.\cite{Liu19} 
Furthermore we note that not only regularly shaped fragments
as considered here host such corner localized states, but more generally
shaped 0D objects can host localized states on the perimeter when
points on opposite sides are related by inversion symmetry. Examples
are shown in SM.\cite{SM}

\paragraph{Robustness against disorder - }  
  It is important to note that the even though our system
  is a Topological Crystal
  Insulator (TCI), which are strictly speaking  only weakly protected, 
  the corner states are fairly robust to bulk disorder as shown in SM.\cite{SM}
  We show this numerically by adding uniformly distributed random on-site bulk terms in a range of the  magnitude of the band gap. 
  This does confirm that approximate symmetries that are preserved on average are enough to host the fractional charges. But as soon as one perturbs the edges, the corner fractionalization gets destroyed. However, because of
  the existence of edge states in the corresponding higher dimensional $1D$
  system,  one still preserves the edge modes in the disordered system
  with perturbed edge states.

\paragraph{Conclusion - }
In this paper we report a physical condensed matter nanoscale realization
of  higher order topological insulator states, namely in the
2D system of buckled monolayer $\beta$-Sb (or other group-V atoms), which has
already been experimentally fabricated although its topological features
have not been reported  yet. By examining
the Wannier centers in this system to be localized on the bond
centers, rather than the atoms, we have shown that this system is a
topological crystal insulator supporting weak topological edge states.
Because of the overall $S_6$ symmetry and its non-zero quadrupole
character, it was shown by a slight generalization
of the symmetry analysis of Ref. \onlinecite{Benalcazar19} to host
topologically protected corner states, similar to those occurring in
the Kekul\'e lattice. 

\paragraph{Acknowledgements -} This work was supported by the U.S. Department of
  Energy-Basic Energy Sciences under grant No. DE-SC0008933.  The calculations made use of the High Performance Computing Resource in the Core Facility for Advanced Research Computing at Case Western Reserve University.

\bibliography{lmto,dft,gw,hoti}

\section{Supplemental Material}

\subsection{Computational Methods}
The band structure calculations of the 2D periodic system 
were performed using density functional theory in the Perdew-Burke-Ernzerhof (PBE)\cite{PBE} generalized   gradient approximation (GGA) both in the
full potential linearized muffin-tin orbital method\cite{questaalpaper}
and using a plane wave
projector augmented wave method (PAW).\cite{PAW}
The band structure of the 2D buckled honeycomb monolayer of
Sb in Fig.2 of the main paper was done 
using the quasiparticle self-consistent
    QS$GW$  many-body perturbation theory method.\cite{Schilfgaarde06,kotani:QSGW}
    Here $GW$ stands
    for the one-electron Green's function and $W$ for the screened
    Coulomb interaction.\cite{Hedin65,Hedin69}
    These calculations were performed
    using the full-potential
    linearized muffin-tin orbital (FP-LMTO) method\cite{Methfessel,Kotani10}
    using the questaal package, which is fully described in Ref.\cite{questaalpaper} and available at \cite{questaal}. Convergence parameters were chosen
    as follows: basis set $spdf-spd$ spherical wave envelope functions  plus augmented plane waves with a cut-off of 3 Ry, augmentation cutoff $l_{max}=4$,
    {\bf k}-point mesh, $12\times12\times2$. The monolayer slabs were
    separated by a vacuum region of 3 nm.  In the $GW$ calculations  the self
    energy $\Sigma$ is calculated  on a {\bf k}-mesh of $5\times5\times2$
    points and interpolated to the above finer mesh and the bands along
    symmetry lines.

    Tight binding calculations were used for the
    nanoribbons or finite size fragments.  We used both a nearest neighbor
    tight-binding model as described in Ref. \onlinecite{radha2019topological}
    or a Wannier interpolation \cite{wannier90}
    of the DFT band structure obtained
    from Quantum Espresso \cite{Giannozzi2009} with kinetic energy  cutoff of 50 Ry.

\begin{figure}[!htb]
  \includegraphics[width=8cm]{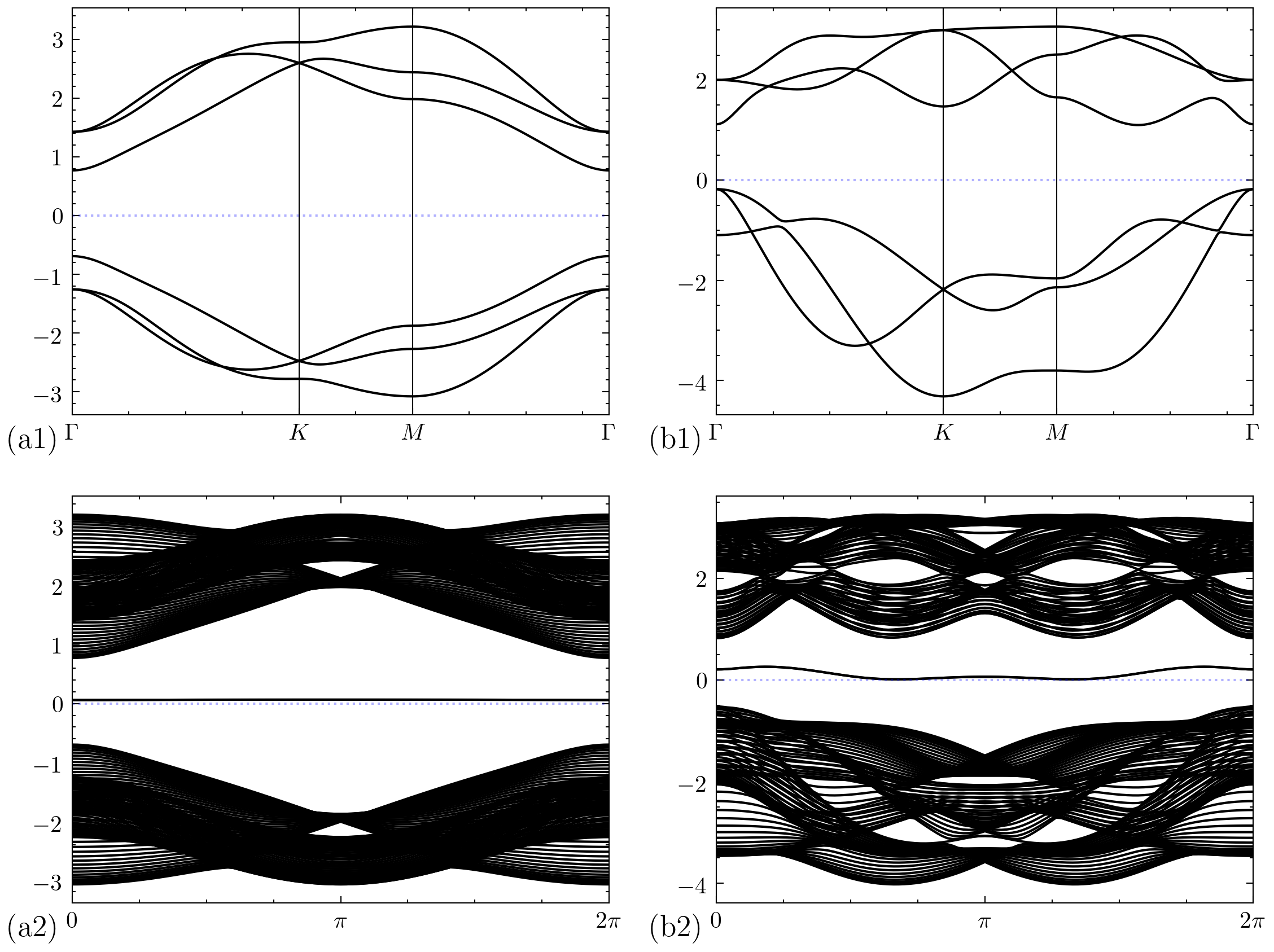} 
  \includegraphics[width=5cm]{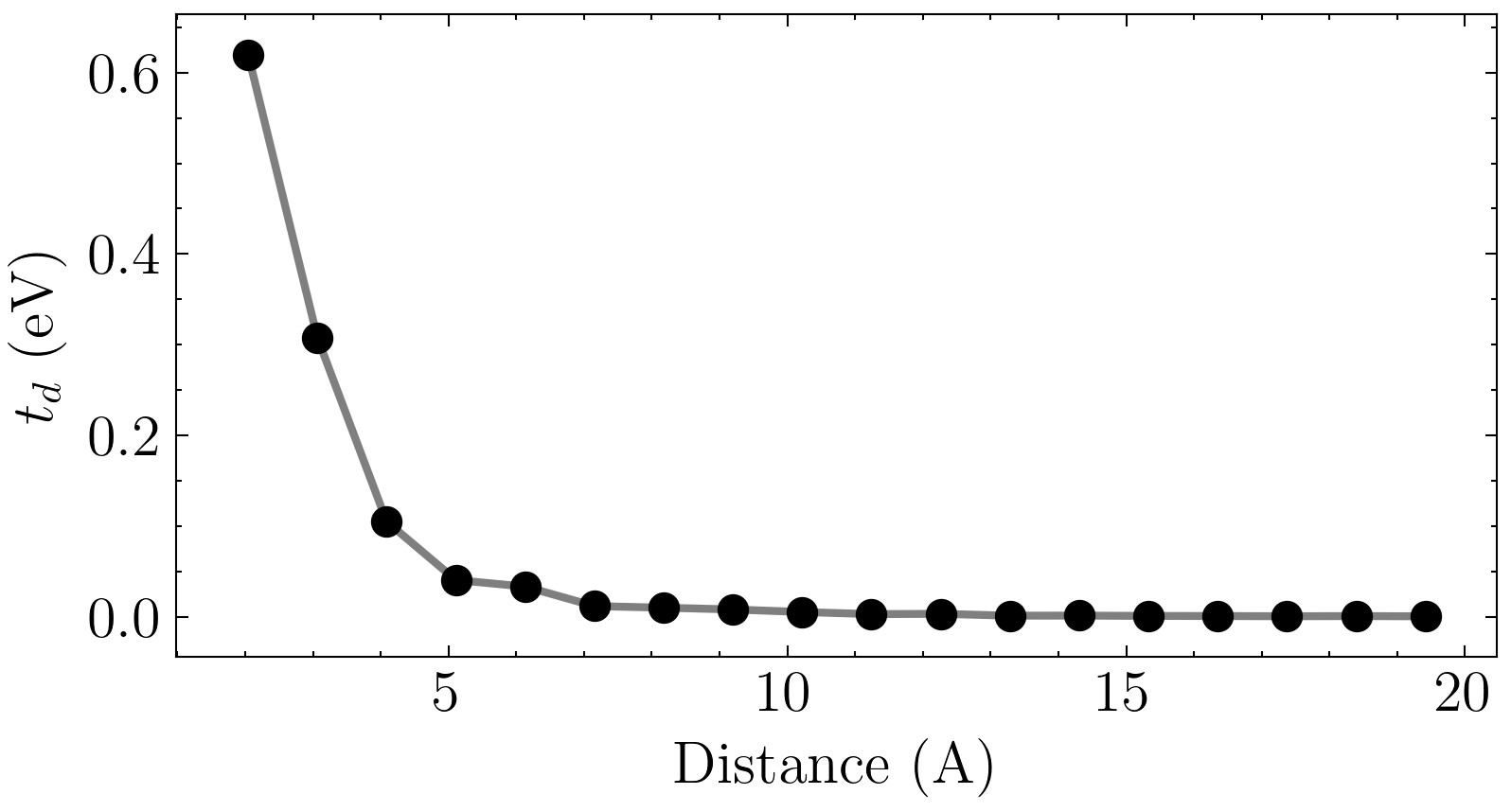}
  \caption{ \textit{(top) } \textit{(a1)} \textit{(a2)} Wannier interpolated LDA bands with $N.N$ interaction and corresponding 1D bands \textit{(b1)} \textit{(b2)} same as above with $N.N.N$. \textit{(bottom) } Wannier interaction strength as function of distance. \label{fig:wannier}}. 
\end{figure}

    \autoref{fig:wannier} shows the 2D and nanoribbon band structures
    obtained within this TB model with either nearest neighbor (NN) or
    next nearest neighbor (NNN) hopping integrals as well as the decay
    of the hopping parameters. It is found that the NNN model already
    reproduces the full DFT plane wave results adequately. The NN model
    is seen to have particle hole symmetry, meaning that the eigenvalues
    occur in pairs $\pm|E_{n{\bf k}}|$. It also shows a completely dispersionless edge band while in the NNN model the edge band shows a small amount of
    dispersion.  The nanoribbon band structure of Fig. 1 in the main
    paper was done with this  NNN Hamiltonian. The
    eigenvalues of the 0D nanofragments were done with the NN TB Hamiltonian
    introduced in Ref. \onlinecite{radha2019topological} which is shown below
    to be close to the NN version
    of this Wannier extracted Hamiltonian.In any case, all these Hamiltonians 
    are topologically equivalent. 
    Tests of the
    main features with NNN TB Hamiltonians for 0D systems were also done
    but do not produced qualitative changes to the conclusions.

\begin{figure*}[!htb]
  \includegraphics[width=16cm]{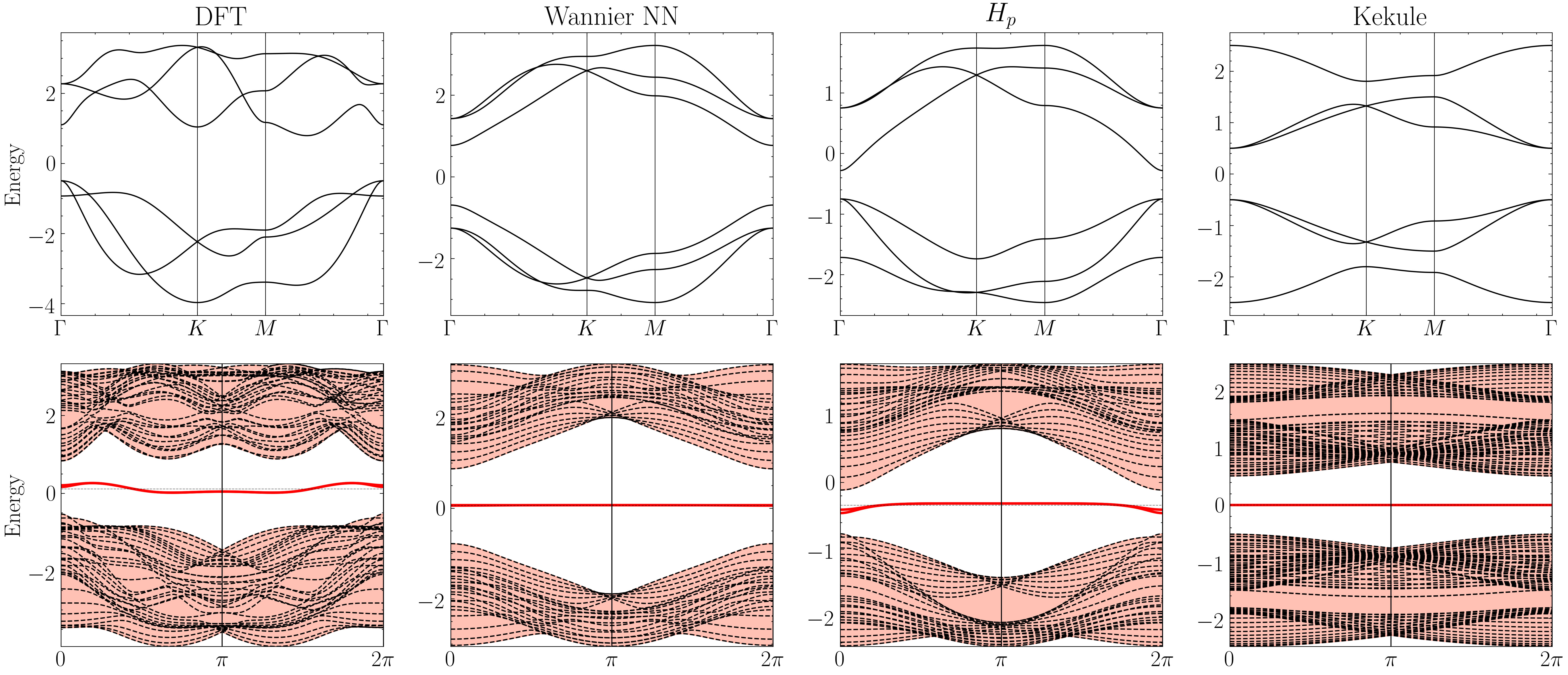} 
  \caption{Bulk and edge band structures at different level of theories\label{fig:bands_compare}}. 
\end{figure*}

In \autoref{fig:bands_compare} we show a further comparison of the DFT, Wannier
NN and two other models. The one labeled $H_p$ corresponds to the 
NN-TB model used in \onlinecite{radha2019topological}. This model
uses $\{p_x,p_y,p_z\}$ orbitals and the standard Slater-Koster
two-center approximation to write the hopping integrals in terms
of $V_{pp\sigma}$ and $V_{pp\pi}$ interactions.  This allowed us to follow
the behavior as function of
buckling angle, assuming the relative amount of $V_{pp\sigma}$ $V_{pp\pi}$
depend only on angle but the bond lengths and hence the $V_{pp\sigma}$
and $V_{pp\pi}$ two-center integrals themselves 
stayed fixed. In this model, also the $p_x,p_y$ orbital energies are slightly
displaced in energy from the $p_z$ orbitals thereby breaking the  particle-hole
symmetry. Although there are changes in the ordering of
bands within the occupied manifold, and within the unoccupied manifold,
the set of irreducible representations comprising the VB and CB manifolds
stay the same and this is the only feature that matters for the topological
aspects discussed in the main paper. 
Finally, we show in this figure the
results of the band structure of the Kekul\'e model, whose tight-binding
Hamiltonian can be written. 

\begin{align}
H=
\setlength{\arraycolsep}{1.4pt}
\renewcommand{\arraystretch}{0.9}
\begin{pmatrix}
0 &\delta & 0 &   e^{i\textbf{k} .\textbf{a}_3} &  0& 1\\ 
\delta & 0 &1  & 0 &  e^{i\textbf{k} .\textbf{a}_2}   & 0\\ 
0& 1&0  & \delta &  0&  e^{i\textbf{k} .\textbf{a}_1}  \\ 
 e^{-i\textbf{k} .\textbf{a}_3} & 0 & \delta & 0 & 1 &0 \\ 
0 &   e^{-i\textbf{k} .\textbf{a}_2} & 0 &1  &  0&  \delta \\ 
1& 0 &   e^{-i\textbf{k} .\textbf{a}_1}  & 0 &  \delta &0
\end{pmatrix} \label{eq:1}
\end{align}
where $\textbf{a}_1=(1,0)$,
$\textbf{a}_2=(\frac{1}{2},\frac{\sqrt{3}}{2})$  The Kekul\'e orbital
centers are numbered 1-6  in Fig.2(c) in the main text and the lattice vectors
are indicated by blue arrows.

\subsection{Kagome and Anti-kagome}
\begin{figure}[!htb]
  \includegraphics[width=9cm]{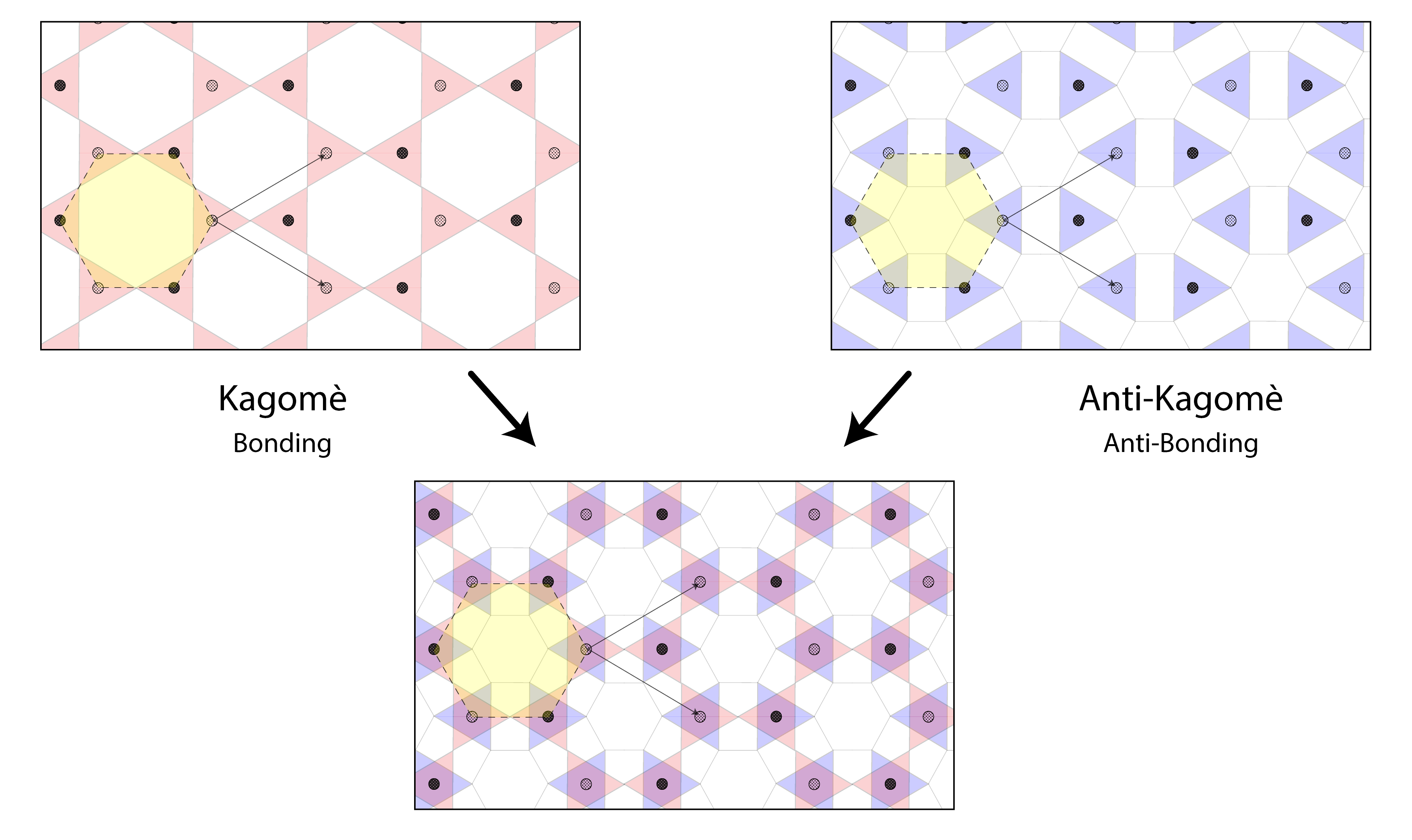} 
  \caption{Bonding and Anti bonding states forming Kagome and Anti kagome\label{fig:kagome}}. 
\end{figure}
\autoref{fig:kagome} show how the bond centers form a Kagome lattice
and the antibond centers form what we call here an anti-Kagome lattice.
The filled and open circles are the Sb atoms point up or down the plane of
the projection plane the corners of the pink triangles are the bond centers
and form a Kagome lattice. On the right, the blue triangle corners are at
the anti-bond centers and form a distinct lattice which we call here
anti-Kagome. Both are shown superposed on each other in the bottom figure.

\subsection{($d-2$) states and symmetry}

\begin{figure}[!htb]
  \includegraphics[width=9cm]{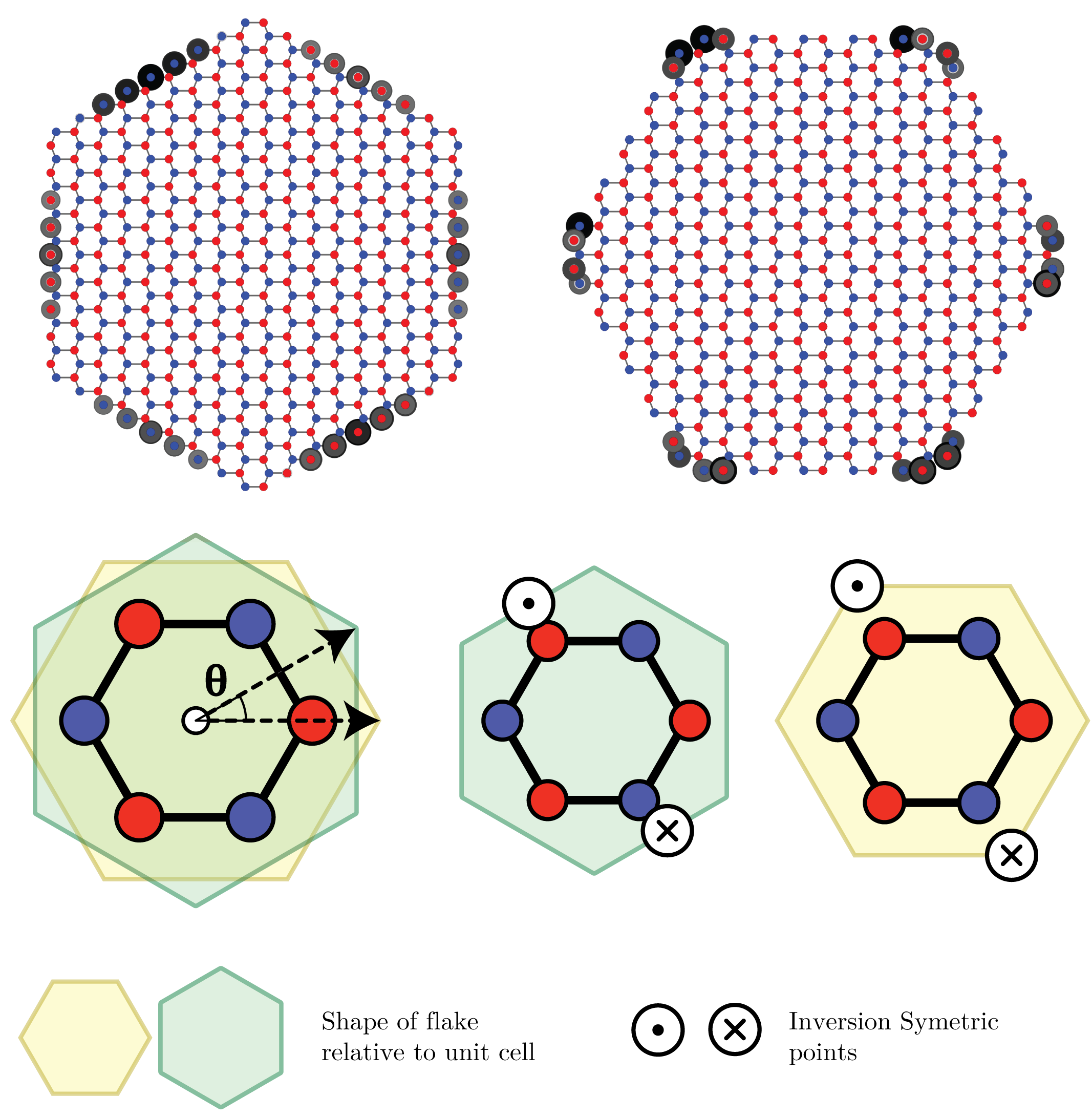} 
  \caption{Different C$_6$ symmetric flake geometry and the mid gap state wavefunction\label{fig:0d-geometry}}. 
\end{figure}
In the main text, we used either a perfectly hexagonal 0D model
or an integer multiple of the 2D primitive unit cell given
by the ${\bf a}_1, {\bf a}_2$ lattice vectors. This lead to symmetrically
distributed corner localized states within each $\pi/6$ sector or a
quadrupole symmetry showing charge localization 
However more generally localized states
at the edges can be obtained from other 
0D fragments as illustrated here in \autoref{fig:0d-geometry}.
These states generally
occur at edge points related to another point on the  circumference
of the 0D system by inversion symmetry 
In the figure on the left we see edge states localized on zigzag edges
but not on the corners joining them, while on the right we see
states localized at the corners joining arm-chair edges. The figures
below illustrate the relation between the overall 0D nanoscale fragment
and the lattice unit cell and which points are related by inversion
which is the key crystallographic symmetry protecting these topological features.

\subsection{Disorder effects}

\begin{figure}[!htb]
  \includegraphics[width=9cm]{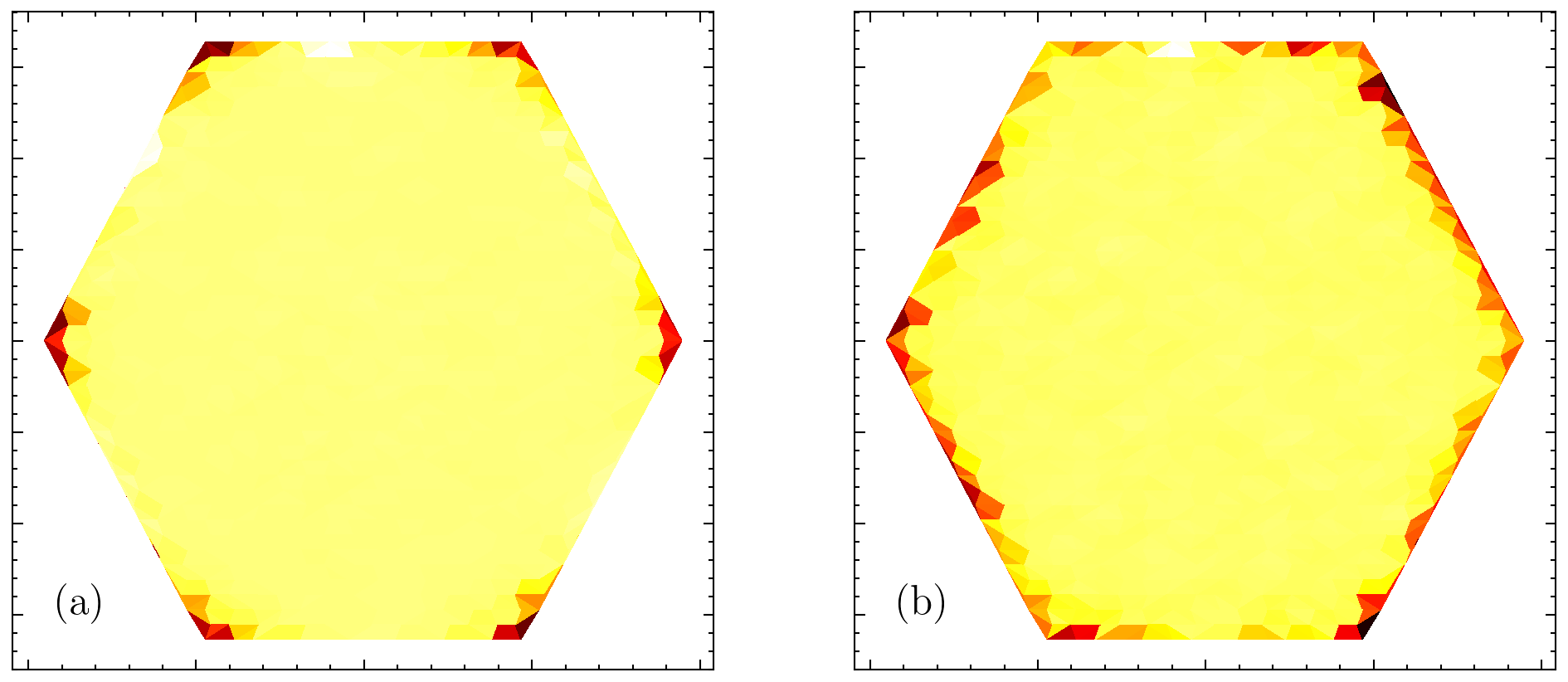} 
  \caption{Total charge distribution of occupied states with on-site potential added stochastically from a uniform distribution of $[-\frac{E_g}{2} ,\frac{E_g}{2}]$   \textit{(a)} everywhere except edge atoms \textit{(b)} everywhere \label{fig:disorder}}. 
\end{figure}

\autoref{fig:disorder} shows the total charge distribution in a $C_6$ symmetric D0 system made up of $4200$ sites $\times 3=12600$ orbitals. Random on-site  potential sampled from a uniform distribution in the interval $[-\frac{E_g}{2} ,\frac{E_g}{2}]$ (where $E_g$ is the band gap) was added on (a) all atoms other than the edges (b) everywhere uniformly. Both simulations were run 500 times and the average charge density is plotted.  This shows that bulk disorder
does not destroy the corner states while disorder also at the edges
does destroy the corner states but still shows localized
states along the edges.

\end{document}